# The Search for >35 MeV Neutrons from the June 3, 2012 Impulsive Flare

K. Koga[1], S. Masuda[2], H. Matsumoto[1], Y. Muraki[2], T. Obara[3]

O. Okudaira[1], S. Shibata[4], T. Yamamoto[5)*], and T. Goka[1]

1) Tsukuba Space Centre, JAXA, Tsukuba 305-8505, Japan
2) Solar-Terrestrial Environment Laboratory, Nagoya University, Nagoya 464-8601, Japan
3) Planetary Plasma and Atmosphere Research Centre, Tohoku University, Sendai 980-8578, Japan
4) Engineering Science Laboratory, Chubu University, Kasugai 487-0027, Japan
5) Department of Physics, Konan University, Kobe 658-8501, japan

*E-mail:* koga.kiyokazu@ jaxa.jp, muraki@stelab.nagoya-u.ac.jp

Abstract

We analyzed a highly impulsive solar flare observed on June 3, 2012. In association with this flare, emissions of hard X-rays, high-energy gamma rays, and neutrons were detected by the detectors onboard the FERMI, RHESSI satellites and the International Space Station. We compared those results with the pictures taken by the UV telescope onboard the Solar Dynamics Observatory satellite and found the crossing structure of two magnetic ropes at two positions on the solar surface almost at the same time. High-energy gamma rays were detected by the Fermi Large Area Telescope satellite, implying that the impulsive flare was one of a major source of proton acceleration processes on the solar surface. At the beginning of research, impulsive solar flares were considered to be the main source of particle acceleration processes; our current observations have confirmed this hypothesis.

1    June 3, 2012 event

On June 3, 2012, a highly impulsive solar flare was observed on the solar surface. The intensity and position of the flare was M3.3 at N16E33. The soft X-ray intensities observed by the Geostationary Operational Environmental Satellite reached a maximum at 17:55 UT. Impulsive emission of hard X-rays was observed by both the Reuven Ramaty High-Energy Solar Spectroscopic Imager (RHESSI) and the Fermi Gamma-ray Burst Monitor (FERMI-GBM) satellites. They recorded very sharp rise and rapid fall-off features. The data from the RHESSI and FERMI-GBM satellites are given in **Figure A**. Furthermore, in association with this flare, high-energy gamma rays were detected by the Fermi Large Area Telescope (FERMI-LAT) satellite. The statistical significance was 8.4 σ, and the intensity was 0.36 (gamma/m$^2$·s > 100 MeV). The rise time of the hard X-rays to the peak was less than 1 min, and the decay time was a few minutes. Understanding of how the particles are accelerated to high energies by such impulsive flares would be an interesting topic. From this viewpoint, we have analyzed the time profiles of hard X-rays, neutrons, and the data from the UV telescope.

We have searched for solar neutrons in the data obtained by the Space Environment Data Acquisition using the FIBer detector (SEDA-FIB) onboard the International Space Station. From 17:50 to 18:30 UT, the ISS flew over the South Pacific Ocean, passing over South America's Cape Horn and the South Atlantic and reaching a point over central Africa. Between 18:13 and 18:20 UT, the ISS flew over the South Atlantic anomaly region, and we avoided analyzing data from this period. The ISS flew in daylight between 17:50 and 18:23 UT. The position of the Sun was on the opposite side of the pressurized module so that separation between solar and background neutrons could be fairly well done.

Solar neutrons were identified in the following way: if the track of the neutron converted proton fell within a cone with an opening angle of less than 30° to 40° from the solar direction, we identified them as being of solar origin. During the flare time, the direction of the Sun was on the opposite side of the pressurized vessel where the cosmonauts were working. Therefore, the solar neutrons and background neutrons were quite easily separated. Furthermore, we used the knowledge that low-energy protons stop at the scintillation bar, leaving a quantity of energy and forming a Bragg peak. The deposited energy was measured by the ADC of the SEDA-FIB. To use these functions, we separated solar neutrons from background neutrons.

We found 26 candidate solar neutrons between 17:54:17 UT and 18:12:48 UT. The observed time and time of departure from the solar surface distributions are shown in **Figure B** by black and red points, respectively. The energy spectrum determined from the arrival time is presented in **Figure C**, separating between two cases, i.e., we assumed two cases for the production time. One of the assumptions was that all the neutrons were simultaneously produced, i.e., at 17:47 UT. The distribution shows the normal production spectrum of solar neutrons; however, here, we encounter a serious issue, i.e., at 17:47 UT, the hard X-ray intensity profile did not start high intensity emission as shown in **Figure A**. Therefore, we assumed the production time of solar neutrons to be near the peak of the hard X-rays or slightly before at 17:53:30 UT. As shown in **Figure D**, the energy spectrum was separated into two components, having low and high energy.

We estimated the background as follows: we selected the data from 90 min before and 90 min after the flare time (17:50–18:30 UT) and searched for neutrons that came from solar direction. For each of the datasets between 16:20 and 16:50 UT and between 19:20 and 19:43 UT, we found three neutrons for which the incident direction coincided with the solar direction. In this case, we avoided data analysis from the time spent passing over the SAA.

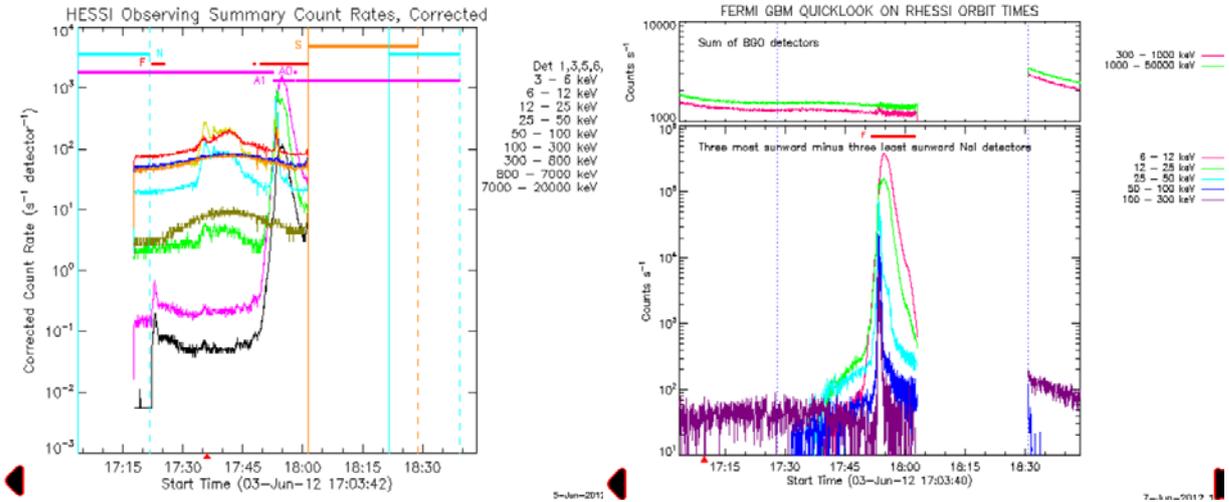

**Figure A**. Time profiles of the observations by the RHESSI and FERMI-GBM satellites during June 3, 2012 (right side). The very fast rise of the intensity can be recognized.

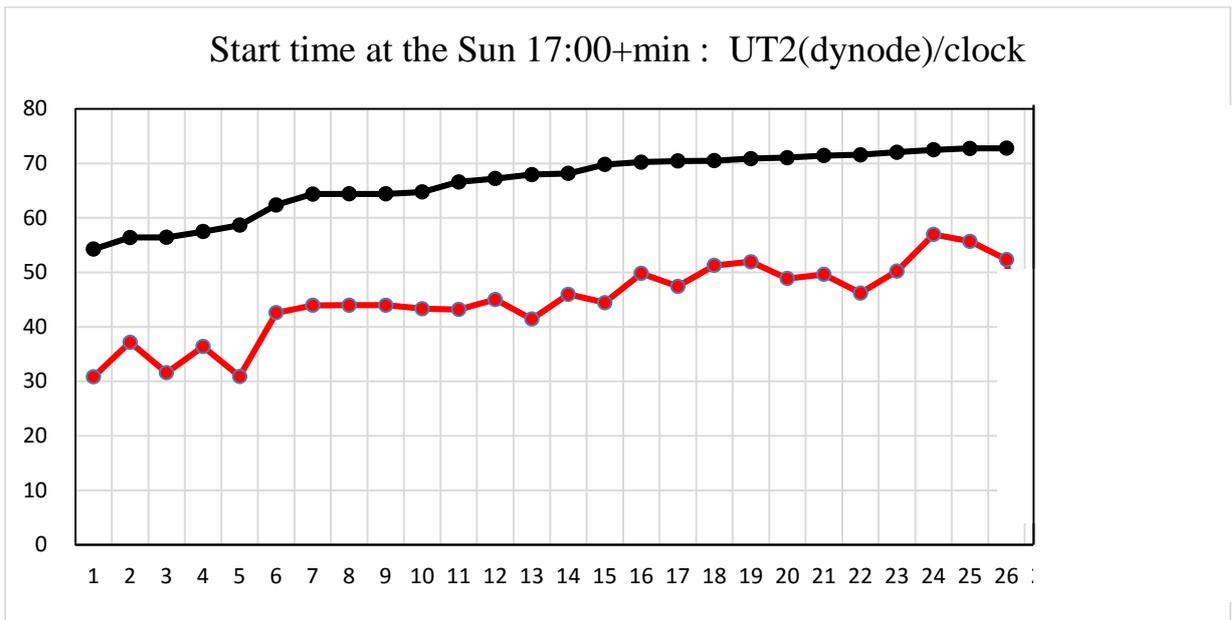

**Figure B**. Arrival time distribution of each event. The ordinate corresponds to the minute after 17:00 UT, and the abscissa represents the number of events. The red points correspond to the production time estimated by the flight time of the neutrons in the inter-planetary space. Actual production time at the Sun must be in between the red and black lines since the neutron detector cannot measure all the energy of the incident neutrons.

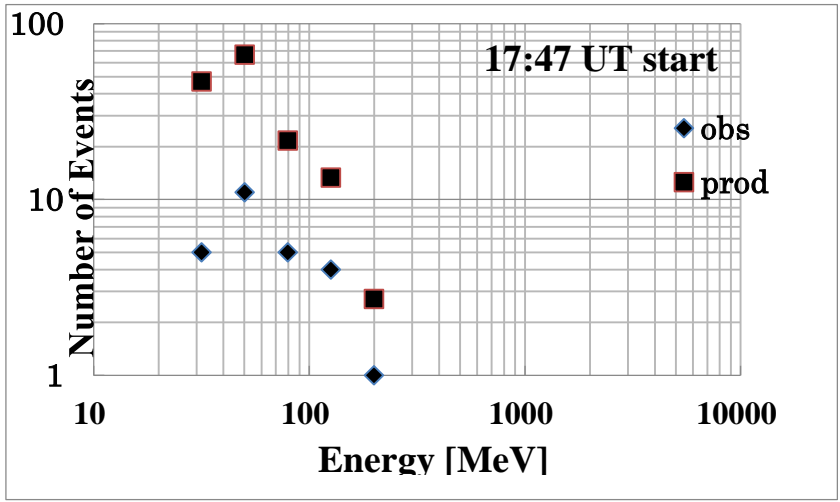

**Figure C.** Energy spectrum of solar neutrons detected on June 3, 2012. The observation value is presented by ♦, whereas the production spectrum is presented by ■. Here, the departure time is assumed to be 17:47:00 UT, i.e., the time estimated by the kinetic energy of the solar neutrons deposited in the sensor.

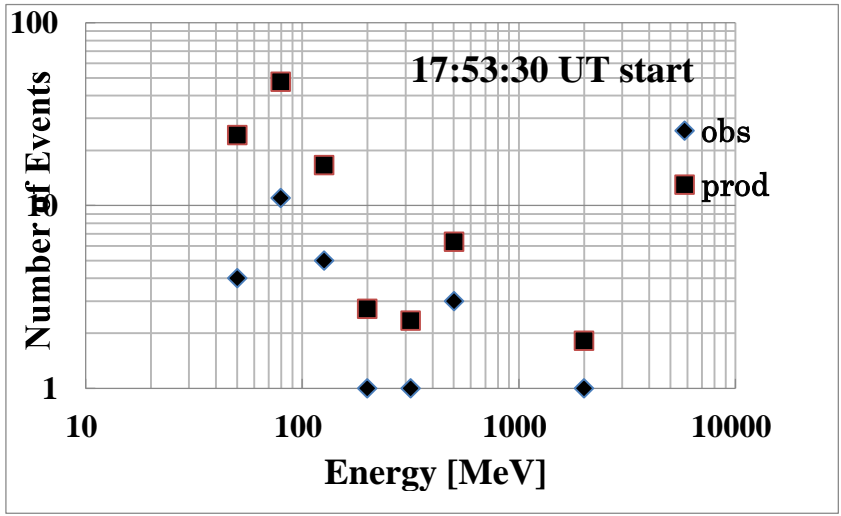

**Figure D.** Energy spectrum of the solar neutrons detected on June 3, 2012. The observation value is presented by ♦ , whereas the production spectrum is presented by ■ . The spectrum seems to have two components, i.e., a high-energy part and a low-energy part (<200 MeV). The production spectrum includes the decay probablility.

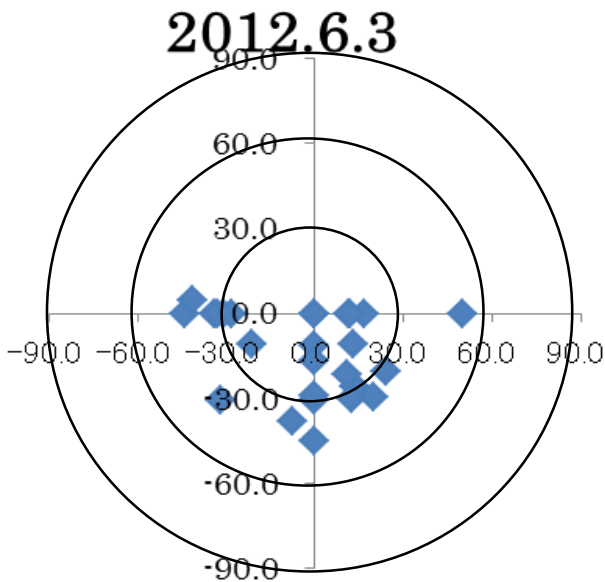

**Figure E**. Arrival direction of solar neutrons measured by using the tracking function of the SEDA-NEM. The direction of the Sun shifts from the lower right-hand side to the lower left-hand side with time.

2      UV pictures obtained by the Solar Dynamical Observatory for June 3, 2012

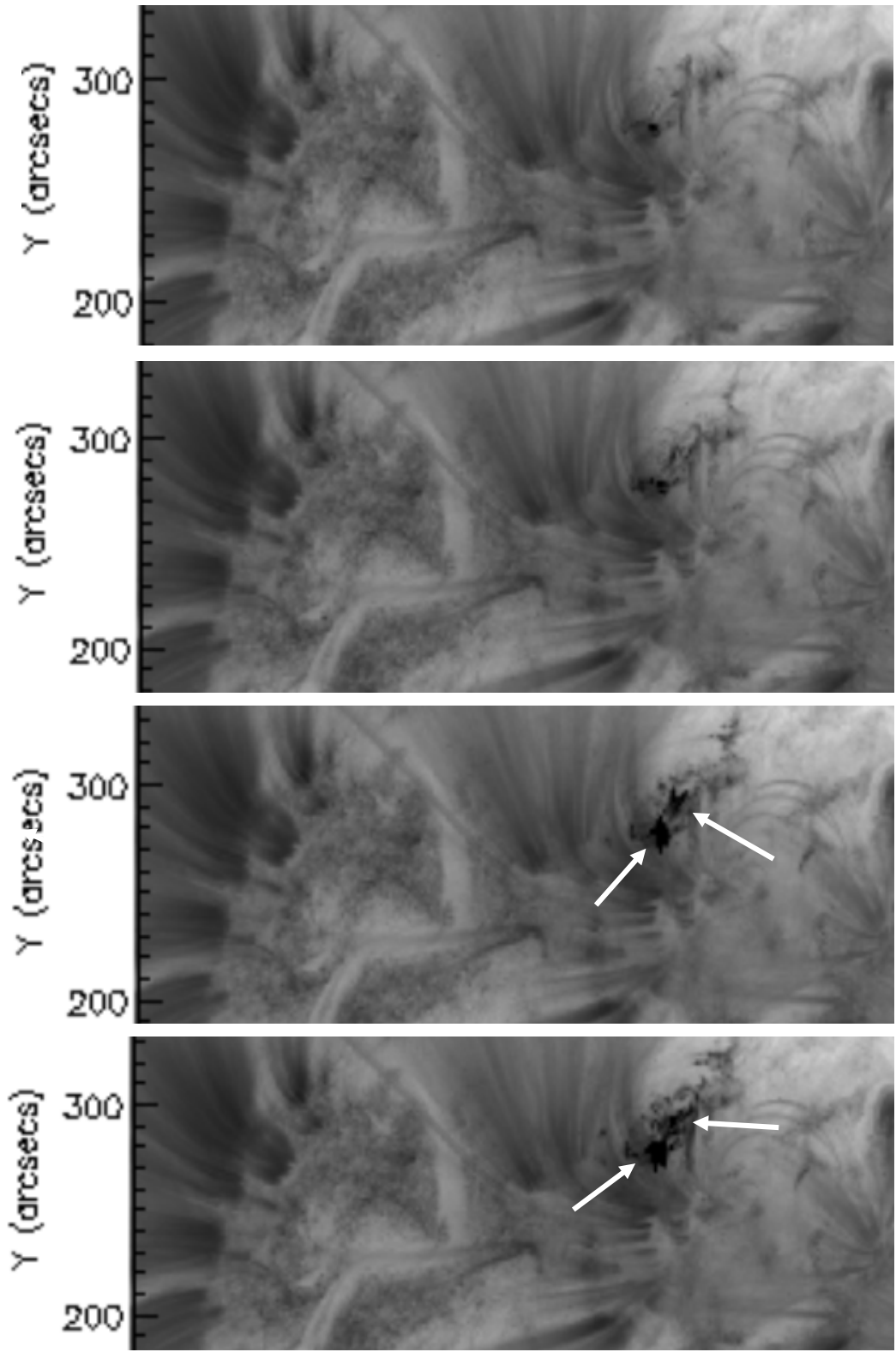

**Figure F**. From top to bottom: Photos taken at 17:51:00, 17:51:24, 17:52:00, and 17:52:24 UT by the 171 nm UV telescope. The white arrows show the positions of the X-type crossing

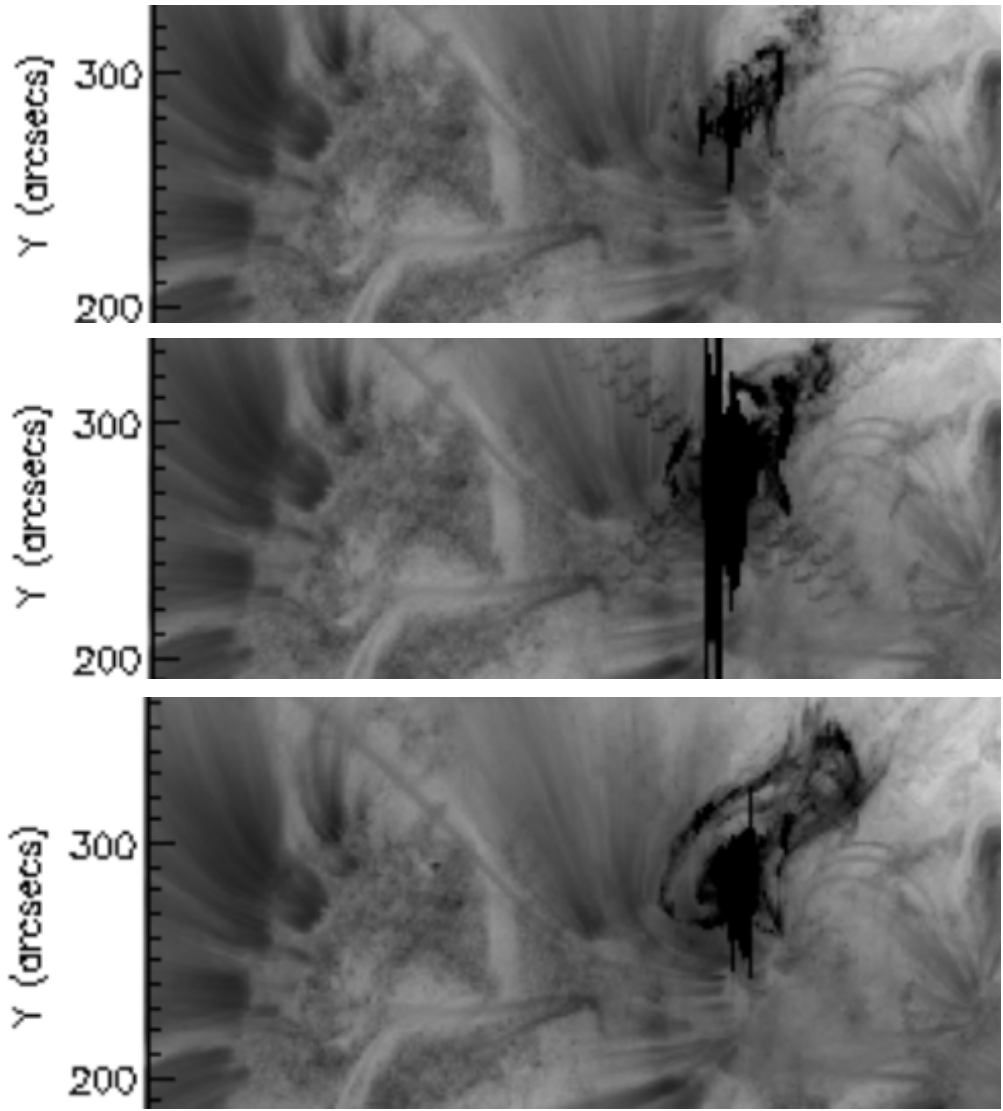

**Figure G.** From top to bottom: Photos taken at 17:52:48, 17:53:13, and 17:54:48 UT using the 171 nm UV telescope of the SDO. Taking account of the pictures presented in Figure F, the flare developed very quickly within a few minutes.

3  Summary of Data Analysis

We have analyzed a highly impulsive solar flare. The emissions of hard X-rays, high-energy gamma rays, and neutrons were detected in association with the flare. When we examined pictures taken by the UV telescope onboard the Solar Dynamics Observatory (SDO) satellite, the crossings of two magnetic rope structures were recognized at two positions and almost at the same time. High-energy gamma rays were detected by the FERMI-LAT satellite, implying that protons were accelerated to high energy (beyond a few GeV) by the impulsive flare. For a long time, impulsive solar flares have been considered to be one of the main sources of particle acceleration processes [1-4]. By current observation with the use of the data detected by many instruments, we have confirmed this hypothesis.

**Acknowledgments**

The authors thank the Tsukuba operation center of Kibo for recording SEDA-NEM-FIB data every day. The authors also acknowledge the NASA Solar Dynamics Observatory team for providing us with valuable data of the UV telescope. We also thank Prof. Gerald Share for valuable discussions on impulsive flares.

# Appendix I   The 131 nm photos taken by the SDO

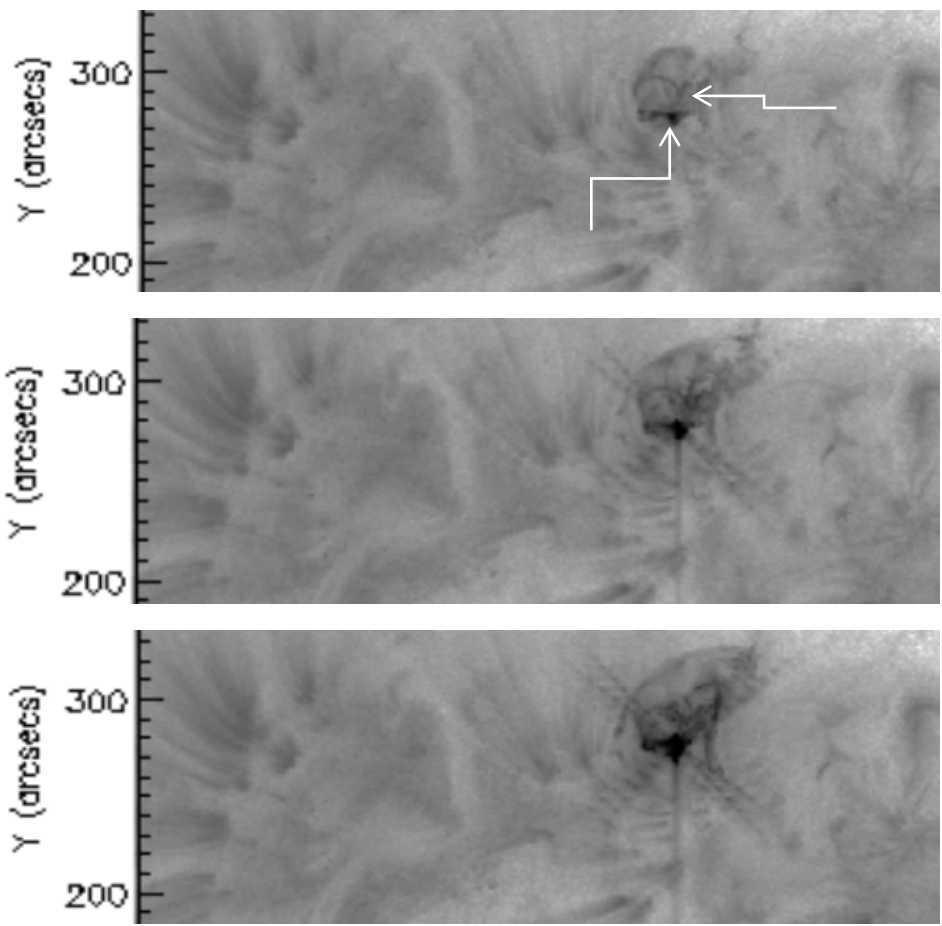

Photos taken at 17:51:09, 17:52:45, and 17:53:09 UT by using the 131 nm SDO UV telescope.

**Appendix II**   Schematic pictures of the classification on the origin of solar flares by Junichi Sakai and Cornelis de Jager

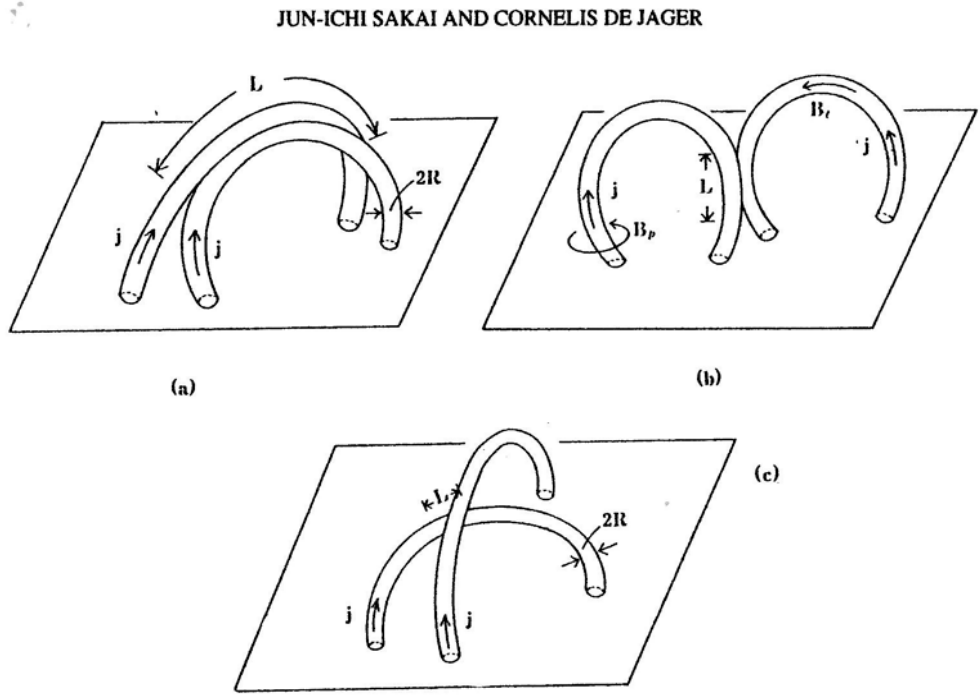

Schematic pictures showing three types of current-loops coalescence: (a) 1-D coalescence (I-type), (b) 2-D coalescence (Y-type), (c) 3-D X-type coalescence. $L$ is the characteristic length of the interacting region. $2R$ is the diameter of the loop with plasma current **J** along the magnetic field $B_t$. $B_p$ is the poloidal magnetic field produced by the plasma current. (Sakai and de Jager, 1991. From *Solar Phys.*)

By courtesy of Junichi Sakai, we cite the pictures from the book entitled "Solar Flares and Collisions between Current-carrying Loops". The same pictures can be also found in the paper of Space Science Review **77** (1996) 1-192. We think that at least the impulsive flares are classified into these patterns. By present observation, we found the type (c) and in another paper by Koga et al (paper #068) we found both the type (b) and the type (c) in the same flare.